# Infeasibility of the nonlocal strain gradient theory for applied Physics

**Mohamed Shaat**∗

*Engineering and Manufacturing Technologies Department, DACC, New Mexico State University, Las Cruces, NM 88003, USA*

*Mechanical Engineering Department, Zagazig University, Zagazig 44511, Egypt*

**Abstract**

In this communication, the feasibility of the nonlocal strain gradient theory for fields of applied mechanics is investigated. It is demonstrated that the nonlocal strain gradient theory is physically incorrect. It is proved that each of the nonlocal theory and the strain gradient theory can model, both, hardening and softening behaviors of materials. Moreover, it is proved that the nonlocal theory and the strain gradient theory describe the same physical phenomena, and hence the strain gradient theory cannot be merged with the nonlocal theory in a unified model. This short communication comments on a series of papers in which the nonlocal strain gradient theory was utilized in different areas of applied mechanics.

**Keywords**: nonlocal theory; strain gradient theory; elasticity; wave propagation; dispersion curves.

## 1. Introduction

Eringen's nonlocal theory has been proposed with the flavor of accounting for the non-neighbor interactions between the building units of the material microstructure [Eringen, 1972a; 1972b; Eringen and Edelen, 1972; Eringen and Kim, 1974; Eringen et al., 1977; Eringen, 1978, 1979]. Eringen's nonlocal theory of linear elasticity emulates the classical theory in considering the infinitesimal strain as the fundamental measure of deformation. However, Eringen's model incorporates an integral operator to sum the nonlocal residuals inside the material; this form is known as integral nonlocal theory [Eringen and Kim, 1974; Eringen et al., 1977; Eringen, 1978, 1979]. In 1983, the integral operator was replaced by a differential operator introducing the differential nonlocal theory [Eringen, 1983].

Eringen's two forms of the nonlocal theory have been used in various studies. Integral nonlocal theory was used to investigate the dispersions of plane waves [Eringen, 1992], stress concentrations near a crack tip [Eringen and Kim, 1974], plasticity and damage of materials [Bazant and Jirasek, 2002], softening plasticity [Jirasek and Rolshoven, 2003], and mechanics of nanostructures [Pisano and Fuschi, 2003; Norouzzadeh and Ansari, 2017; Fernández-Sáez et al., 2016; Tuna and Kirca, 2016a; Tuna and Kirca, 2016b]. Uncountable list of studies utilized the differential nonlocal theory in investigating the mechanics of nanostructures [*e.g.* Lu et al., 2006; Reddy, 2007; Shaat, 2015; Lu et al., 2007; Zhang et al., 2017; Liew et al., 2017].

---

∗Corresponding author. Tel.: +15756215929.
 *E-mail addresses:* shaat@nmsu.edu; shaatscience@yahoo.com (M. Shaat).

Another group of nonclassical theories is the strain gradient and couple stress theories. Mindlin [Mindlin 1964, 1965; Mindlin and Eshel, 1968] has proposed the strain gradient theory for linear elastic continua in which the strain energy depends on the infinitesimal strain as well as its gradients. Moreover, the couple stress theory [Mindlin and Tierestan, 1962; Toupin, 1962] has been developed such that the internal energy depends on the strain and the first-order rotation gradient. An extension of the couple stress theory by developing the second rotation gradient theory was proposed by incorporating the 2nd-order rotation gradient [Shaat and Abdelkefi, 2016].

In fact, the field of applied nonlocal mechanics suffers from misleading information. For example, it is thought that nonlocal models only account for softening behaviors of materials. The strain gradient and the couple stress theory, on the contrary, only account for hardening behaviors of materials. Thus, it is thought that the nonlocal parameter of the nonlocal theory and the length scales of the strain gradient and couple stress theories describe different material characteristics [*e.g.* Lim et al., 2015]. This motivated some researchers [*e.g.* Lim et al., 2015; Attia and Mahmoud, 2016] to combine the nonlocal theory with the strain gradient and/or the couple stress theory in a unified model. It was thought that these unified models give the flavor of combining the softening and hardening mechanisms. However, these attempts should be revised for their feasibilities.

It was thought that Eringen's nonlocal model of linear elasticity can be generalized by the attempt proposed by Lim et al. [2015] by developing a nonlocal strain gradient theory in which the strain energy function depends on two independent integral operators for the infinitesimal strain and its first-order gradient. In fact, this attempt needs to be revised for its feasibility. Lim, Zhang and Reddy [2015] claimed, "*nonlocal elastic models can only account for softening stiffness with [an] increasing nanoscale parameter.*" Then, they added, "*the stiffness enhancement effect, noticed from experimental observation[s] and as well as the gradient elasticity (or modified couple stress) theories cannot be characterized,*" by the nonlocal theory, they meant. Moreover, they mentioned, "*Many related research works have been performed to model the static and dynamic problems based on various gradient theories and all the results showed a stiffness enhancement effect with an increase of the gradient coefficients.*" Accordingly, they concluded, "*the length scales present in Erinegn's nonlocal elasticity model and the strain gradient theories describe two entirely different physical characteristics of materials and structures at nanoscale.*" Therefore, they literally mentioned, as the main motivation behind their proposed theory, "*There is a need to bring both of the length scales into a single theory so that the true effect of the two length scales on the structural response can be assessed.*"

It follows from these quotations that the nonlocal strain gradient theory was proposed with the aim of developing a unified model that can account for the softening and hardening behaviors of materials. However, in fact, the hardening and the softening behaviors of materials can be captured by either the nonlocal theory or the strain gradient theory and no need to merge them in a single model. In this study, a proof for this fact is presented.

In this communication, the fact that each of the nonlocal elasticity theory and the strain gradient theory can reflect, both, softening and hardening behaviors of materials is demonstrated. Moreover, it is revealed that these two theories describe the same physical phenomena and cannot be merged in a unified theory. This argues against the claims by Lim et al. [2015] and Attia and Mahmoud [2016] and demolishes the concept of the nonlocal strain gradient.

## 2. The nonlocal theory models softening and hardening behaviors

Referring to Eringen's book, "Nonlocal continuum field theories," [Eringen, 2002, p. 99], the nondimensional wave frequency, $\bar{\omega}/\bar{c}_0$, can be written in the context of the nonlocal elasticity theory as follows:

$$\frac{\bar{\omega}}{\bar{c}_0} = \bar{k}\sqrt{\frac{1}{1 + \beta\bar{k}^2 + \delta\bar{k}^4}} \qquad (1)$$

where $\bar{k}$ denotes the nondimensional wave number. $\beta$ and $\delta$ are two nondimensional nonlocal parameters.

When $\delta = 0$ and $\beta = \tau^2$, the dispersion relation (1) recovers the one presented in Lim et al. [2015] for Eringen's nonlocal theory. As stated by Eringen [2002, p. 99], for better fitting with the dispersion curves of materials, $\beta$ and $\delta$ may possess negative or positive values. However, researchers have confined the use of these nonlocal parameters to $\delta = 0$ and $\beta = \tau^2 > 0$. In fact, utilizing these two parameters and considering their possible negative and positive values, the nonlocal theory can reflect, both, hardening and softening behaviors of materials, as shown in Figure 1. The nonlocal theory can reflect the softening behavior (the decrease in the nondimensional wave frequency) when $\beta > 0$ and $\delta = 0$, $\beta \geq 0$ and $\delta > 0$, or $\delta + \beta \geq 0$ and $\delta < 0$. On the other hand, the nonlocal theory depicts the hardening behavior (the increase in the nondimensional wave frequency) when $\beta < 0$ and $\delta = 0$, $\delta + \beta \leq 0$ and $\delta > 0$, or $\beta \leq 0$ and $\delta < 0$. Indeed, Figure 1 demonstrates that the nonlocal theory can give all the possible configurations of the dispersion curves. This contradicts with the claims that the nonlocal theory can only reflect softening behaviors of materials and nanostructures.

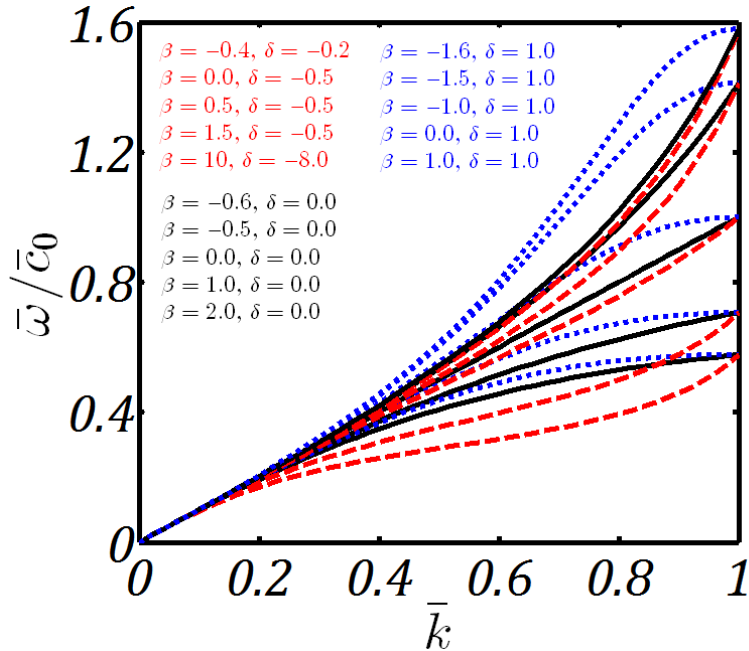

Figure 1: The nonlocal theory can reflect softening and hardening behaviors of materials. This theory depicts all the possible configurations of dispersion curves.

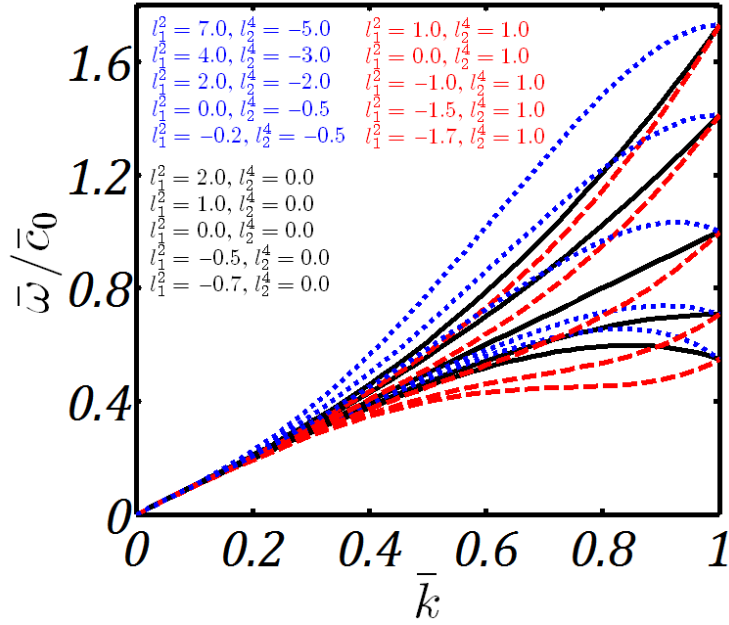

Figure 2: The strain gradient theory can reflect softening and hardening behaviors of materials. This theory depicts all the possible configurations of dispersion curves.

### 3. The strain gradient theory models softening and hardening behaviors

Referring to Mindlin [1965] and Shaat [2017], the dispersion relation can be written according to the strain gradient theory (up to the second-order strain gradient) in the form:

$$\frac{\overline{\omega}}{\overline{c}_0} = \overline{k}\sqrt{1 + l_1^2 \overline{k}^2 + l_2^4 \overline{k}^4} \qquad (2)$$

where $l_1$ and $l_2$ are two nondimensional length scales. When $l_2 = 0$ and $l_1 = \xi$, the dispersion relation (2) recovers the one presented in Lim et al. [2015] for the strain gradient theory.

It should be mentioned that Mindlin has clearly stated that the length scales may possess real or complex values (*i.e.* $l_1^2, l_2^4 \geq 0$ or $l_1^2, l_2^4 \leq 0$) [Mindlin, 1965, p. 424; Mindlin and Tiersten, 1962, p. 416]. However, the use of these length scales in the literature was only confined for the real values. Shaat [2017] demonstrated that the real and complex forms of these length scales are needed in order to give the strain gradient theory the ability to model the softening and hardening behaviors of materials and nanostructures. As shown in Figure 2, utilizing all the possible forms of the length scales, the strain gradient theory can reflect both hardening and softening behaviors of materials and nanostructures. The strain gradient theory reflects a softening behavior when $l_1^2 < 0$ and $l_2^4 = 0$, $l_1^2 + l_2^4 \leq 0$ and $l_2^4 > 0$, or $l_1^2 \leq 0$ and $l_2^4 < 0$. However, the strain gradient theory gives a hardening behavior when $l_1^2 > 0$ and $l_2^4 = 0$, $l_1^2 \geq 0$ and $l_2^4 > 0$, or $l_1^2 + l_2^4 \geq 0$ and $l_2^4 < 0$. As depicted in Figure 2, the strain gradient theory can give all the possible configurations of the dispersion curves. This, as well, argues against the claims that the strain gradient theory can only give the hardening behaviors of materials and nanostructures.

## 4. The strain gradient theory is a nonlocal theory

To demonstrate the fact that the nonlocal theory and the strain gradient theory describe the same physical phenomena, different dispersion curves as obtained by these two theories are plotted in Figure 3. As shown in the figure, the strain gradient theory can exactly fit the nonlocal theory when the length scales, $l_1$ and $l_2$, are properly related to the nonlocal parameters, $\beta$ and $\delta$, [Shaat, 2017]. Indeed, the two theories can reflect the dispersive nature of the acoustic dispersion curves of materials, as shown in Figure 3. This demonstrates the fact that the strain gradient theory is a nonlocal theory where it accounts for the non-neighbor interactions exactly the same as the nonlocal theory [Shaat, 2017]. Hence, the strain gradient theory cannot be merged with the nonlocal theory in a unified model.

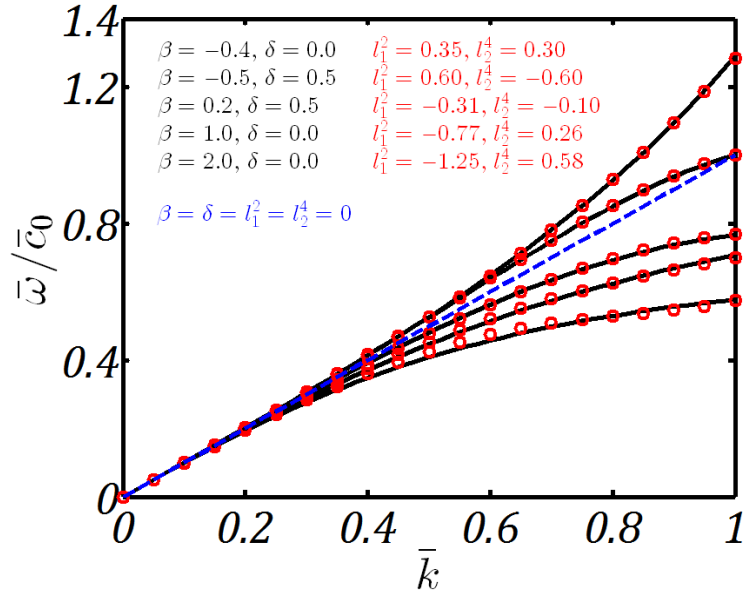

Figure 3: The nonlocal theory and the strain gradient theory describe the same physical phenomena. These two theories give the same dispersion curves. (*Dispersions as obtained by the nonlocal theory are presented by dark solid curves while the dispersions as obtained by the strain gradient theory are depicted by open circles*).

## 5. Conclusions

It follows from the derived results and the presented discussions that the nonlocal strain gradient theory [Lim et al., 2015] and the nonlocal couple stress theory [Attia and Mahmoud, 2016] are physically incorrect. It was demonstrated that the strain gradient theory can present the same phenomena exactly the same as the nonlocal theory. Moreover, both theories can reflect the softening and hardening behaviors of materials.

The nonlocal strain gradient theory has been utilized in various studies including [*e.g.* Li et al., 2015; 2016a; 2016b; Li and Hu, 2015; Ebrahimi and Barati, 2016; 2017; Farajpour et al., 2016]. Although of the wide spread of this theory among researchers, it is not physically possible to combine the nonlocal theory with the strain gradient theory nor the couple stress theory in unified models. Therefore, these studies should be revised for their feasibility.

**References**

Attia, M.A., Mahmoud, F.F., 2016. Modeling and analysis of nanobeams based on nonlocal couple-stress elasticity and surface energy theories. *International Journal of Mechanical Sciences* 105, 126–134.

Bazant, Z.P., Jirasek, M., 2002. Nonlocal Integral Formulations of Plasticity and Damage: Survey of Progress. *Journal of Engineering Mechanics*, 128(11), 1119-1149.

Ebrahimi, F. Barati, M. R., 2016. Wave propagation analysis of quasi-3D FG nanobeams in thermal environment based on nonlocal strain gradient theory. *Applied Physics A* 122, 843.

Ebrahimi, F., Barati, M., 2017. Hygrothermal effects on vibration characteristics of viscoelastic FG nanobeams based on nonlocal strain gradient theory. *Composite Structures* 159, 433-444.

Eringen, A.C., 1972a. Nonlocal polar elastic continua. *Int. J. Eng. Sci.*, 10, 1–16.

Eringen, A.C., 1972b. Linear theory of nonlocal elasticity and dispersion of plane waves. *Int. J. Eng. Sci.*, 10, 425–435.

Eringen, A.C., Edelen, D.G.B., 1972. On nonlocal elasticity. *Int J Eng Sci*, 10(3), 233–48.

Eringen, A.C., Kim, B.S., 1974. Stress concentration at the tip of a crack. *Mechanics Research Communications*, 1, 233–237.

Eringen, A.C., Speziale, C.G., Kim, B.S., 1977. Crack-tip problem in non-local elasticity. *Journal of Mechanics Physics and Solids*, 25, 339–355.

Eringen, A.C., 1978. Line crack subjected to shear. *International Journal of Fracture*, 14, 367–379.

Eringen, A.C., 1979. Line crack subjected to anti-plane shear. *Engineering Fracture Mechanics*, 12, 211–219.

Eringen, A. C., 1983. On differential equations of nonlocal elasticity and solutions of screw dislocation and surface waves. *J. Appl. Phys.*, 54 (9), 4703-4710.

Eringen, A.C., 1992. Vistas of nonlocal continuum physics. *Int J Eng Sci*, 30(10), 1551–65.

Eringen AC. Nonlocal continuum field theories. NewYork: Springer-Verlag; 2002.

Fernández-Sáez, J. , Zaera, R. , Loya, J. A. , Reddy, J. N., 2016. Bending of Euler–Bernoulli beams using Eringen's integral formulation: A paradox resolved. *International Journal of Engineering Science*, 99, 107–116.

Farajpour, A, Haeri Yazdi, M. R., Rastgoo, A., Mohammadi, M., 2016. A higher-order nonlocal strain gradient plate model for buckling of orthotropic nanoplates in thermal environment. *Acta Mechanica* 227(7), 1849–1867.

Jirasek, M., Rolshoven, S., 2003. Comparison of integral-type nonlocal plasticity models for strain-softening materials. *International Journal of Engineering Science*, 41, 1553–1602.

Lu, P., Lee, H.P., Lu, C., 2006. Dynamic properties of flexural beams using a nonlocal elasticity model. *J Appl Phys*, 99, 073510.

Li, L., Hu, Y., Ling, L., 2015. Flexural wave propagation in small-scaled functionally graded beams via a nonlocal strain gradient theory. *Composite Structures* 133, 1079–1092.

Li, L., Li, X., Hu, Y. 2016a. Free vibration analysis of nonlocal strain gradient beams made of functionally graded material. *International Journal of Engineering Science* 102, 77–92.

Li, L., Hu, Y., 2015. Buckling analysis of size-dependent nonlinear beams based on a nonlocal strain gradient theory. International Journal of Engineering Science 97 (2015) 84–94.

Li, L., Hu, Y., Ling, L., 2016b. Wave propagation in viscoelastic single-walled carbon nanotubes with surface effect under magnetic field based on nonlocal strain gradient theory. *Physica E* 75, 118–124.

Liew, K.M., Zhang, Y., Zhang, L. W., 2017. Nonlocal elasticity theory for graphene modeling and simulation: prospects and challenges. Journal of Modeling in Mechanics and Materials, 1(1), doi:10.1515/jmmm-2016-0159.

Lim, C.W., Zhang, G., Reddy, J.N., 2015. A Higher-order nonlocal elasticity and strain gradient theory and its Applications in wave propagation. *Journal of the Mechanics and Physics of Solids*. 78, 298–313.

Lu, P., Zhang, P.Q., Lee, H.P., Wang, C.M., Reddy, J.N., 2007. Non-local elastic plate theories. *Proc R Soc A*, 463, 3225–40.

Mindlin, R.D., 1964. Microstructure in linear elasticity, *Arch. Rational Mech. Anal.* 16, 51–78.

Mindlin, R.D., 1965. Second gradient of strain and surface-tension in linear elasticity. *International Journal of Solids and Structure* 1, 417–438.

Mindlin, R. D. and Tiersten, H. F., 1962. Effects of couple stresses in linear elasticity, *Arch. Rational Mech. Anal.* 11, 415–448.

Mindlin, R.D., Eshel, N.N., 1968. On first strain-gradient theories in linear elasticity, *Int. J. Solids Struct.* 4, 109-124.

Norouzzadeh, A., Ansari, R., 2017. Finite element analysis of nano-scale Timoshenko beams using the integral model of nonlocal elasticity. *Physica E*, 88, 194–200.

Pisano, A.A., Fuschi, P., 2003. Closed form solution for a nonlocal elastic bar in tension. Int J Solids Struct, 41, 13–23.

Reddy, J.N., 2007. Nonlocal theories for bending, buckling and vibration of beams. *Int J Eng Sci*, 45, 288–307.

Shaat, M., 2015. Iterative nonlocal elasticity for Kirchhoff plates. *Int J Mech Sci*, 90, 162–70.

Shaat, M. and Abdelkefi, A., 2016. On a second-order rotation gradient theory for linear elastic continua, *International Journal of Engineering Science* 100, 74–98.

Shaat M., 2017. A general nonlocal theory and its approximations for slowly varying acoustic waves. *Int J Mech Sci* 130, 52–63.

Tuna, M., Kirca, M., 2016a. Exact solution of Eringen's nonlocal integral model for bending of Euler-Bernoulli and Timoshenko beams. *International Journal of Engineering Science* 105, 80–92.

Tuna, M., Kirca, M., 2016b. Exact solution of Eringen's nonlocal integral model for vibration and buckling of Euler-Bernoulli beam. *International Journal of Engineering Science* 107, 54-67.

Toupin, R.A., 1962. Elastic materials with couple-stresses, *Arch. Rational Mech. Anal.* 11, 385–414.

Zhang, L.W., Zhang, Y., Liew, K.M., 2017. Vibration analysis of quadrilateral graphene sheets subjected to an inplane magnetic field based on nonlocal elasticity theory. *Composites Part B*, 118, 96-103.